\documentclass[aps,superscriptaddress,reprint]{revtex4-1}
\usepackage{amsmath}
\usepackage{amssymb}
\usepackage{graphicx}
\usepackage[colorlinks,citecolor=blue, linkcolor=blue,hyperindex,CJKbookmarks,dvipdfm]{hyperref}

\begin{document}

\title{Enantiomeric excess determination based on nonreciprocal transition induced spectral line elimination}
\author{Xun-Wei Xu}
\email{davidxu0816@163.com}
\affiliation{Department of Applied Physics, East China Jiaotong University, Nanchang,
330013, China}
\author{Chong Ye}
\affiliation{Beijing Computational Science Research Center, Beijing 100193,
China}
\author{Yong Li}
\email{liyong@csrc.ac.cn}
\affiliation{Beijing Computational Science Research Center, Beijing 100193,
China} \affiliation{Synergetic Innovation Center for Quantum Effects and
Applications, Hunan Normal University, Changsha 410081, China}
\author{Ai-Xi Chen}
\email{aixichen@zstu.edu.cn}
\affiliation{Department of Physics, Zhejiang Sci-Tech University, Hangzhou
310018, China}
\affiliation{Department of Applied Physics, East China Jiaotong University, Nanchang,
330013, China}
\date{\today }

\begin{abstract}
The spontaneous emission spectrum of a multi-level atom or molecule with nonreciprocal
transition is investigated. It is shown that the nonreciprocal transition
can lead to the elimination of a spectral line in the spontaneous emission
spectrum. As an application, we show that nonreciprocal transition arises from the phase-related driving
fields in chiral molecules with cyclic three-level transitions, and the
elimination of a spectral line induced by nonreciprocal transition provides
us a method to determine the enantiomeric excess for the chiral molecules without requiring the enantio-pure
samples.
\end{abstract}

\maketitle

\section{Introduction}

Nonreciprocity is a very general concept that arises in many branches of
physics, such as electronics, optics~\cite{JalasNPT13,CalozPRAPP18},
acoustics~\cite{MaznevaWM13,FleurySci14}, and condensed-matter physics~\cite{AtalaNPy14}.
Nonreciprocity means that some particles or waves exhibit different
transmission properties when their sources and detectors are exchanged, such
as the one-way electric conduction in the semiconductor p-n junctions. In a
recent paper~\cite{XWXuArx19}, some of us introduced the concept of nonreciprocity to
investigate the transitions between different energy levels, and proposed a
generic method to realize significant difference between the stimulated
emission and absorption coefficients of two nondegenerate energy levels,
which we refer to as nonreciprocal transition~\cite{XWXuArx19}. The
nonreciprocal transition can be used for many applications, such as
single-photon nonreciprocal transporter~\cite{XWXuArx19}, nonreciprocal phonon devices~\cite{JZhangPRB15,YJiangPRAPP18}, and
the echo cancellation in quantum memory~\cite{LvovskyNPo09} and quantum measurements~\cite{ClerkRMP10}.

In this paper, we will study the spontaneous emission of a multi-level atom or molecule with
nonreciprocal transition. The spontaneous emission spectrum emitted from an
atom or molecule shows a valuable insight into the behaviour of transitions
between different energy levels~\cite{SYZhuPRA95,SYZhuPRL96}. We find that
the nonreciprocal transition can be reflected with spectral line elimination
in the spontaneous emission spectra, which provides us a very simple way to
test nonreciprocal transition in experiment.

On the other hand, chirality is important in chemistry and biology for many
chemical and biological processes are chirality-dependent~\cite{BuschEls06}. But the chiral
discrimination and separation remain challenging works under the existing experimental
technical conditions. Some spectroscopic methods have been developed to
determine enantiomeric excess, such as circular dichroism~\cite%
{StephensJPC85,FanoodNC15}, Raman optical activity~\cite{YHeAS11,BegzjavOE19}, and spectroscopy for a cyclic three-level model~\cite{AnsariPRA90,YXLiuPRL05,CYePRA18} based on quantum interference effects~\cite{WZJiaPRA11}, three-wave mixing ~\cite{HirotaPJASB12,PattersonNat13,PattersonPRL13,PattersonPCCP14,ShubertACIE14,LobsigerJPCL15,ShubertJPCL15,ShubertJCP15,EibenbergerPRL17,PerezACIE17,LeibscherJCP19,KochRMP19}%
, ac Stark effect~\cite{CYePRA19}, or deflection effect~\cite{YYChenArx19}.
In addition, many methods were proposed to achieve inner-state separating~\cite%
{KralPRL01,YLiPRA08,WZJiaJPB10,LehmannJCP18,VitanovPRL19,CYePRA19,CYearX19,JLWuarX19} or spatially separating~\cite{YLiPRL07,XLiJCP10,JacobJCP12,BrandPRL18,MilnerPRL19,SuzukiPRA19} molecules of different chiralities, and
enantioconversion of chiral mixtures~\cite{ShapiroPRL00,BrumerPRA01,GerbasiJCP01,KralPRL03,FrishmanJPB04,CYeArx19}.

In the latter part of this paper, we apply the general theory of the
spontaneous emission of multi-level systems to the chiral molecules. We show
that nonreciprocal transition arises from the phase-related driving fields
in chiral molecules with cyclic three-level transitions and the appearance
of spectral line elimination in the spontaneous emission spectra around some
resonant frequencies is chirality-dependent. Therefore, the enantiomeric excess can
be determined by measuring the spontaneous emission spectra of chiral molecules.
Different from
the traditional methods of enantiomeric excess determination~\cite{StephensJPC85,FanoodNC15,YHeAS11}, the
spectral line elimination in the spontaneous emission spectrum induced by
nonreciprocal transition provides us another method to determine the
enantiomeric excess for chiral molecules without requiring the enantio-pure
samples.

The remainder of this paper is organized as follows. In Sec.~II, the basic
theory for spontaneous emission spectrum of a general multi-level system with
nonreciprocal transition is introduced. The time evolution of the
populations and the corresponding spontaneous emission spectra of the
multi-level system with nonreciprocal transition are investigated in detail in
Sec.~III. The application of the spontaneous emission spectra in determining
the enantiomeric excess for chiral molecules are discussed in Sec.~IV.
Finally, a summary is given in Sec.~V.

\section{Basic theory for spontaneous emission spectrum}

\begin{figure}[tbp]
\includegraphics[bb=4 376 586 613, width=8.5 cm, clip]{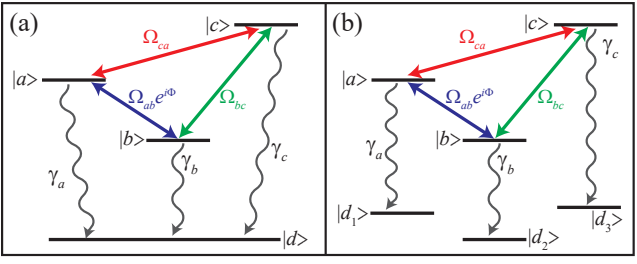}
\caption{(Color online) Level diagram of an atom or molecule with cyclic transitions for
the three upper levels ($|a\rangle$, $|b\rangle$ and $|c\rangle$), and they
are coupled by the same vacuum modes to (a) the common lower level ($%
|d\rangle$), or (b) different lower levels ($|d_1\rangle$, $|d_2\rangle$ and
$|d_3\rangle$).}
\label{fig1}
\end{figure}

We study the spontaneous emission of an atom or molecule with three upper levels ($%
|a\rangle$, $|b\rangle$, and $|c\rangle$), which are coupled to each other by three
strong fields with frequencies ($\nu_{ab}$, $\nu_{cb}$, and $\nu_{ca}$), Rabi
frequencies ($\Omega_{ab}$, $\Omega_{cb}$, and $\Omega_{ca}$) and phases ($%
\phi_{ab}$, $\phi_{cb}$, and $\phi_{ca}$). The spontaneous emission spectrum
for the system will be derived for two different cases: (A) the three upper
levels are coupled to one common lower level ($|d\rangle$) with the same
vacuum modes, as shown in Fig.~\ref{fig1}(a), or (B) the three upper levels
are coupled to three different lower levels ($|d_1\rangle$, $|d_2\rangle$,
and $|d_3\rangle$) respectively with the same vacuum modes, as shown in Fig.~%
\ref{fig1}(b).

\begin{widetext}
\subsection{With one common lower level}

Consider the three upper levels ($|a\rangle $, $|b\rangle $, and $|c\rangle $%
) coupled with one common lower level ($|d\rangle $) by the same vacuum
modes. The interaction Hamiltonian of the system in the interaction picture
can be written as ($\hbar =1$)
\begin{eqnarray}
V &=&\Omega _{ab}e^{i\Phi }e^{i\Delta _{ab}t}\left\vert a\right\rangle
\left\langle b\right\vert +\Omega _{cb}e^{i\Delta _{cb}t}\left\vert
c\right\rangle \left\langle b\right\vert +\Omega _{ca}e^{i\Delta
_{ca}t}\left\vert c\right\rangle \left\langle a\right\vert  \notag
\label{Eq1} \\
&&+\sum_{k}\left[ g_{k}^{ad}e^{i\left( \omega _{ad}-\omega _{k}\right)
t}v_{k}\left\vert a\right\rangle \left\langle d\right\vert
+g_{k}^{bd}e^{i\left( \omega _{bd}-\omega _{k}\right) t}v_{k}\left\vert
b\right\rangle \left\langle d\right\vert +g_{k}^{cd}e^{i\left( \omega
_{cd}-\omega _{k}\right) t}v_{k}\left\vert c\right\rangle \left\langle
d\right\vert \right.  \notag \\
&&+\mathrm{H.c.},
\end{eqnarray}%
where $\omega _{\sigma \sigma^{\prime }}$ $\left( \sigma,\sigma^{\prime
}=a,b,c,d\right) $ are the frequency differences between levels $|\sigma
\rangle $ and $| \sigma ^{\prime }\rangle $, $\Delta _{\sigma \sigma^{\prime
}}=\omega _{\sigma \sigma^{\prime }}-\nu _{\sigma \sigma^{\prime }}$ is the
detuning of the driving fields, $v_{k}$ ($v_{k}^{\dagger }$) is the
annihilation (creation) operator for the $k$th vacuum mode with frequency $%
\omega _{k}$, and $g_{k}^{\sigma d}$ is the coupling constant between the $k$%
th vacuum mode and the atomic transition from $|\sigma \rangle $ to $%
|d\rangle $. Here real $g_{k}^{\sigma d}$ is assumed, $k$ denotes both the momentum and polarization of the
vacuum modes, and the total phase $%
\Phi =\phi _{ab}-\phi _{cb}+\phi _{ca}$ is obtained by redefining $e^{-i\phi
_{cb}}|b\rangle \rightarrow |b\rangle $ and $e^{-i\phi _{ca}}|a\rangle
\rightarrow |a\rangle $.

We assume the system is initially prepared in one of the upper levels, i.e., $|\psi(0)\rangle=|b\rangle|0\rangle$, where $|0\rangle$ denotes the vacuum state. The state
vector at time $t$ can be written as%
\begin{equation}
\left\vert \psi \left( t\right) \right\rangle =\left[ A\left( t\right)
\left\vert a\right\rangle +B\left( t\right) \left\vert b\right\rangle
+C\left( t\right) \left\vert c\right\rangle +\sum_{k}D_{k}\left( t\right)
v^\dag_{k}\left\vert d\right\rangle\right] \left\vert 0\right\rangle ,
\end{equation}%
where the modulus squares of the coefficients $A(t)$, $B(t)$, $C(t)$ and $D_{k}(t)$ are the occupation
probabilities in the corresponding state at time $t$. By using the
Weisskopf-Wigner approximation, the dynamical behaviors for the coefficients
are given by%
\begin{eqnarray}  \label{Eq3}
\frac{d}{dt}A\left( t\right) &=&-\frac{\gamma _{a}}{2}A\left( t\right)
-p_{ab}\frac{\sqrt{\gamma _{a}\gamma _{b}}}{2}e^{i\omega _{ab}t}B\left(
t\right) -p_{ca}\frac{\sqrt{\gamma _{a}\gamma _{c}}}{2}e^{-i\omega
_{ca}t}C\left( t\right)  \notag \\
&&-i\Omega _{ab}e^{i\Phi }e^{i\Delta _{ab}t}B\left( t\right) -i\Omega
_{ca}e^{-i\Delta _{ca}t}C\left( t\right),
\end{eqnarray}%
\begin{eqnarray}  \label{Eq4}
\frac{d}{dt}B\left( t\right) &=&-\frac{\gamma _{b}}{2}B\left( t\right)
-p_{ab}\frac{\sqrt{\gamma _{a}\gamma _{b}}}{2}e^{-i\omega _{ab}t}A\left(
t\right) -p_{cb}\frac{\sqrt{\gamma _{b}\gamma _{c}}}{2}e^{-i\omega
_{cb}t}C\left( t\right)  \notag \\
&&-i\Omega _{bc}e^{-i\Delta _{cb}t}C\left( t\right) -i\Omega _{ab}e^{-i\Phi
}e^{-i\Delta _{ab}t}A\left( t\right),
\end{eqnarray}%
\begin{eqnarray}  \label{Eq5}
\frac{d}{dt}C\left( t\right) &=&-\frac{\gamma _{c}}{2}C\left( t\right)
-p_{ca}\frac{\sqrt{\gamma _{a}\gamma _{c}}}{2}e^{i\omega _{ca}t}A\left(
t\right) -p_{cb}\frac{\sqrt{\gamma _{c}\gamma _{b}}}{2}e^{i\omega
_{cb}t}B\left( t\right)  \notag \\
&&-i\Omega _{ca}e^{i\Delta _{ca}t}A\left( t\right) -i\Omega _{bc}e^{i\Delta
_{cb}t}B\left( t\right),
\end{eqnarray}%
\begin{equation}  \label{Eq6}
\frac{d}{dt}D_{k}\left( t\right) =-ig_{k}^{ad}e^{-i\left( \omega
_{ad}-\omega _{k}\right) t}A\left( t\right) -ig_{k}^{bd}e^{-i\left( \omega
_{bd}-\omega _{k}\right) t}B\left( t\right) -ig_{k}^{cd}e^{-i\left( \omega
_{cd}-\omega _{k}\right) t}C\left( t\right),
\end{equation}
where $\gamma_a=[2\pi
(g^{ad}_k)^2\rho(\omega_k)]_{\omega_k=\omega_{ad}}$, $\gamma_b=[2\pi
(g_k^{bd})^2\rho(\omega_k)]_{\omega_k=\omega_{bd}}$ and $\gamma_c=[2\pi(
g_k^{cd})^2\rho(\omega_k)]_{\omega_k=\omega_{cd}}$ are the decay rates, $%
\rho(\omega_k)$ is the mode density, $p_{\sigma\sigma^{\prime }}=\vec{\mu}%
_\sigma\cdot \vec{\mu}_{\sigma^{\prime }}/(|\vec{\mu}_\sigma|\cdot|\vec{\mu}%
_{\sigma^{\prime }}|)$ and $\vec{\mu}_\sigma$ ($\vec{\mu}_{\sigma^{\prime }}$%
) is the the dipole moment of the transition from $|\sigma\rangle$ ($%
|\sigma^{\prime }\rangle$) to $|d\rangle$ ($\sigma,\sigma^{\prime }=a,b,c$).
The coupling terms with $p_{\sigma \sigma^{\prime }}$ is induced by the
decays from different upper levels ($|\sigma \rangle$ and $|\sigma'\rangle$) to the common lower level $|d\rangle$, which can result in spontaneous emission cancellation and spectral line elimination~\cite%
{SYZhuPRA95,SYZhuPRL96}. However, here $\{\omega _{ab},\omega _{ca},\omega
_{cb}\}\gg \{\gamma_a,\gamma_b,\gamma_c,\Delta _{ab},\Delta
_{ca},\Delta_{cb}\}$ are assumed so that the high-frequency oscillating
terms (i.e., the decay induced coupling terms with $p_{\sigma\sigma^{\prime }}\frac{\sqrt{\gamma _{\sigma}\gamma _{\sigma^{\prime }}}}{2}e^{\pm i\omega_{\sigma\sigma^{\prime }}t}$) in Eqs.~(\ref{Eq3})-(\ref{Eq5}%
) can be neglected, and the dynamical equations are simplified as
\begin{equation}  \label{Eq7}
\frac{d}{dt}A\left( t\right) =-\frac{\gamma _{a}}{2}A\left( t\right)
-i\Omega _{ab}e^{i\Phi }e^{i\Delta _{ab}t}B\left( t\right) -i\Omega
_{ca}e^{-i\Delta _{ca}t}C\left( t\right),
\end{equation}%
\begin{equation}  \label{Eq8}
\frac{d}{dt}B\left( t\right) =-\frac{\gamma _{b}}{2}B\left( t\right)
-i\Omega _{bc}e^{-i\Delta _{cb}t}C\left( t\right) -i\Omega _{ab}e^{-i\Phi
}e^{-i\Delta _{ab}t}A\left( t\right),
\end{equation}%
\begin{equation}  \label{Eq9}
\frac{d}{dt}C\left( t\right) =-\frac{\gamma _{c}}{2}C\left( t\right)
-i\Omega _{ca}e^{i\Delta _{ca}t}A\left( t\right) -i\Omega _{bc}e^{i\Delta
_{cb}t}B\left( t\right).
\end{equation}
For convenience of calculations, let us define $\widetilde{B}\left( t\right) \equiv
e^{i\Delta _{ab}t}B\left( t\right) $ and $\widetilde{C}\left( t\right)
\equiv e^{-i\Delta _{ca}t}C\left( t\right) $ and make the assumption of three-photon resonance $\Delta
_{ca}+\Delta _{ab}=\Delta _{cb}$, then we get a dynamical equations with constant coefficients as
\begin{equation}  \label{Eq10}
\frac{dA\left( t\right) }{dt}=-\frac{\gamma _{a}}{2}A\left( t\right)
-i\Omega _{ab}e^{i\Phi }\widetilde{B}\left( t\right) -i\Omega _{ca}%
\widetilde{C}\left( t\right),
\end{equation}%
\begin{equation}  \label{Eq11}
\frac{d\widetilde{B}\left( t\right) }{dt}=-\left( \frac{\gamma _{b}}{2}%
-i\Delta _{ab}\right) \widetilde{B}\left( t\right) -i\Omega _{bc}\widetilde{C%
}\left( t\right) -i\Omega _{ab}e^{-i\Phi }A\left( t\right),
\end{equation}%
\begin{equation}  \label{Eq12}
\frac{d\widetilde{C}\left( t\right) }{dt}=-\left( \frac{\gamma _{c}}{2}%
+i\Delta _{ca}\right) \widetilde{C}\left( t\right) -i\Omega _{ca}A\left(
t\right) -i\Omega _{bc}\widetilde{B}\left( t\right).
\end{equation}%
All the following calculations and discussions on the occupation
probabilities are based on Eqs.~(\ref{Eq10})-(\ref{Eq12}).

In the following, we will use the Laplace transform method to solve the
dynamic equations. By taking the Laplace transform, i.e., $\overline{O}\left( s\right)
=\int_{0}^{+\infty }O\left( t\right) e^{-st}dt$, of Eqs.~(\ref{Eq10})-(\ref%
{Eq12}), with the initial condition $\Psi _{0}=\left[ A\left( 0\right) ,%
\widetilde{B}\left( 0\right) ,\widetilde{C}\left( 0\right) \right] ^{T}$, we
get
\begin{equation}  \label{Eq13}
\overline{\Psi }=M^{-1}\Psi_0
\end{equation}%
with $\overline{\Psi }=\left[ \overline{A}\left( s\right) ,\overline{%
\widetilde{B}}\left( s\right) ,\overline{\widetilde{C}}\left( s\right) %
\right] ^{T}$, and
\begin{equation}
M=\left(
\begin{array}{ccc}
s+\frac{\gamma _{a}}{2} & i\Omega _{ab}e^{i\Phi } & i\Omega _{ca} \\
i\Omega _{ab}e^{-i\Phi } & s+\frac{\gamma _{b}}{2}-i\Delta _{ab} & i\Omega
_{bc} \\
i\Omega _{ca} & i\Omega _{bc} & s+\frac{\gamma _{c}}{2}+i\Delta _{ca}%
\end{array}%
\right).
\end{equation}

The spontaneous emission spectrum of the system, $S\left( \omega \right) $, is
the Fourier transform of%
\begin{eqnarray}
\left\langle E^{-}\left( t+\tau \right) E^{+}\left( t\right) \right\rangle
_{t\rightarrow +\infty } &\equiv &\left\langle \psi \right\vert
\sum_{k,k^{\prime }}v_{k}^{\dag }e^{i\omega _{k}\left( t+\tau \right)
}v_{k^{\prime }}e^{-i\omega _{k^{\prime }}t}\left\vert \psi \right\rangle
_{t\rightarrow +\infty }  \notag \\
&=&\int_{-\infty }^{+\infty }\left\vert D_{k}\left( +\infty \right)
\right\vert ^{2}\rho \left( \omega _{k}\right) e^{i\omega _{k}\tau }d\omega
_{k},
\end{eqnarray}%
then we have $S\left( \omega _{k}\right) =\left\vert D_{k}\left(
+\infty\right) \right\vert ^{2}\rho \left( \omega _{k}\right) $, where $%
D_{k}\left( +\infty\right)$ is the long time behavior ($t\rightarrow +\infty $) of $D_{k}\left(
t\right)$ and can be obtained by integrating time $t^{\prime }$ in Eq.~(\ref{Eq6}) as
\begin{eqnarray}
D_{k}\left( +\infty \right) &=&\int_{0}^{ +\infty }\left[ -ig_{k}^{ad}e^{-i%
\left( \omega _{ad}-\omega _{k}\right) t^{\prime }}A\left( t^{\prime
}\right) -ig_{k}^{bd}e^{-i\left( \omega _{bd}-\omega _{k}+\Delta
_{ab}\right) t^{\prime }}\widetilde{B}\left( t^{\prime }\right)
-ig_{k}^{cd}e^{-i\left( \omega _{cd}-\omega _{k}-\Delta _{ca}\right)
t^{\prime }}\widetilde{C}\left( t^{\prime }\right) \right] dt^{\prime } \nonumber\\
&=&-i g_{k}^{ad}\overline{A}\left( -i\delta
_{k}\right) -ig_{k}^{bd}\overline{\widetilde{B}}\left( -i \delta^b _{k}
\right) -ig_{k}^{cd}\overline{\widetilde{C}}\left( -i\delta^c _{k}\right)
\end{eqnarray}
with the detunings $\delta _{k}\equiv \omega _{k}-\omega _{ad}$, $\delta^b
_{k}\equiv \delta _{k}+\omega _{ab}-\Delta _{ab} $, $\delta^c _{k}\equiv
\delta _{k}-\omega _{ca}+\Delta _{ca}$. Finally, the spontaneous emission
spectrum is given by
\begin{equation}  \label{Eq17}
S\left( \omega _{k}\right) =\frac{1}{2\pi }\left\vert \sqrt{\gamma _{a}}%
\overline{A}\left( -i\delta _{k}\right) +\sqrt{\gamma _{b}}\overline{%
\widetilde{B}}\left( -i \delta^b _{k} \right) +\sqrt{\gamma _{c}}\overline{%
\widetilde{C}}\left( -i\delta^c _{k}\right) \right\vert ^{2},
\end{equation}
with $\overline{A}\left( s\right)$, $\overline{\widetilde{B}}\left( s\right)$%
, and $\overline{\widetilde{C}}\left( s\right)$ given by Eq.~(\ref{Eq13}).

\subsection{With three different lower levels}

In the case with three different lower levels, the interaction Hamiltonian of the system is
given in the interaction picture by
\begin{eqnarray}
V^{\prime } &=&\Omega _{ab}e^{i\Phi }e^{i\Delta _{ab}t}\left\vert
a\right\rangle \left\langle b\right\vert +\Omega _{bc}e^{i\Delta
_{cb}t}\left\vert c\right\rangle \left\langle b\right\vert +\Omega
_{ca}e^{i\Delta _{ca}t}\left\vert c\right\rangle \left\langle a\right\vert
\notag \\
&&+\sum_{k}\left[ g_{k}^{ad_1}e^{i\left( \omega _{ad_1}-\omega _{k}\right) t}v_{k}\left\vert a\right\rangle \left\langle
d_{1}\right\vert +g_{k}^{bd_2}e^{i\left( \omega _{bd_2}-\omega _{k}\right) t}v_{k}\left\vert b\right\rangle \left\langle
d_{2}\right\vert +g_{k}^{cd_3}e^{i\left( \omega _{cd_3}-\omega _{k}\right) t}v_{k}\left\vert c\right\rangle \left\langle
d_{3}\right\vert \right]   \notag \\
&&+\mathrm{H.c.},
\end{eqnarray}%
where $\omega _{\sigma d_j}$ denote the frequency differences
between levels $|\sigma \rangle $ $\left( \sigma =a,b,c\right) $ and $%
|d_{j}\rangle $ ($j=1,2,3$), and $g_{k}^{\sigma d_j}$ is the
coupling constant between the $k$th vacuum mode and the atomic transition
from $|\sigma \rangle $ to $|d_{j}\rangle $. The state vector for the system
at time $t$ can be written as%
\begin{equation}
\left\vert \psi^{\prime } \left( t\right) \right\rangle=\left[ A\left(
t\right) \left\vert a\right\rangle +B\left( t\right) \left\vert
b\right\rangle +C\left( t\right) \left\vert c\right\rangle \right]
\left\vert 0\right\rangle +\sum_{j=1,2,3}\sum_{k}D_{j,k}\left( t\right)
\left\vert d_{j}\right\rangle \left\vert 1\right\rangle _{k}.
\end{equation}%
By substituting the Hamiltonian and state vector into the Schr\"{o}dinger equation and using the Weisskopf-Wigner approximation, the dynamic equations of the coefficients [$A(t)$, $B(t)$, and $C(t)$] are obtained and they have the same forms as those given by Eqs.~(\ref{Eq7})-(\ref{Eq9}), with the decay rates replaced by $\gamma^{\prime }_a=[2\pi
(g^{ad_1}_k)^2\rho(\omega_k)]_{\omega_k=\omega_{ad_1}}$, $\gamma^{\prime }_b=[2\pi
(g_k^{bd_2})^2\rho(\omega_k)]_{\omega_k=\omega_{bd_2}}$, and $\gamma^{\prime }_c=[2\pi(
g_k^{cd_3})^2\rho(\omega_k)]_{\omega_k=\omega_{cd_3}}$. The spontaneous emission spectrum is obtained as $S\left( \omega
_{k}\right) =\sum_{j=1,2,3}\left\vert D_{j,k}\left( +\infty \right)
\right\vert ^{2}\rho \left( \omega _{k}\right) $, where
\begin{equation}
D_{1,k}\left( +\infty \right) =-ig_{k}^{ad_1}\overline{A}\left(
-i\delta _{k}^{\prime }\right) ,
\end{equation}%
\begin{equation}
D_{2,k}\left( +\infty \right) =-ig_{k}^{bd_2}\overline{\widetilde{B}}%
\left( -i\delta _{k}^{b\prime }\right) ,
\end{equation}%
\begin{equation}
D_{3,k}\left( +\infty \right) =-ig_{k}^{cd_3}\overline{\widetilde{C}}%
\left( -i\delta _{k}^{c\prime }\right) ,
\end{equation}%
the coefficients [$\overline{A}\left( s\right) $, $\overline{\widetilde{B}}%
\left( s\right) $, and $\overline{\widetilde{C}}\left( s\right) $] are given
by Eq.~(\ref{Eq13}), and the detunings are defined by $\delta _{k}^{\prime
}\equiv \omega _{k}-\omega _{ad_1}$, $\delta _{k}^{b\prime }\equiv
\omega _{k}-\omega _{bd_2}-\Delta _{ab}=\delta _{k}^{\prime }+\omega
_{ab}^{\prime }-\Delta _{ab}$, $\delta _{k}^{c\prime }\equiv \omega
_{k}-\omega _{cd_3}+\Delta _{ca}=\delta _{k}^{\prime }-\omega
_{ca}^{\prime }+\Delta _{ca}$, with $\omega _{ab}^{\prime }\equiv \omega
_{ad_1}-\omega _{bd_2}$, $\omega _{ca}^{\prime }\equiv
\omega _{cd_3}-\omega _{ad_1}$. Thus, the spontaneous
emission spectrum for the case with three different lower levels can be
specifically expressed as
\begin{equation}
S\left( \omega _{k}\right) =\frac{1}{2\pi }\left\{ \gamma _{a}\left\vert
\overline{A}\left( -i\delta _{k}^{\prime }\right) \right\vert ^{2}+\gamma
_{b}\left\vert \overline{\widetilde{B}}\left( -i\delta _{k}^{b\prime
}\right) \right\vert ^{2}+\gamma _{c}\left\vert \overline{\widetilde{C}}%
\left( -i\delta _{k}^{c\prime }\right) \right\vert ^{2}\right\} .
\end{equation}
For comparison with the case of one common lower level given by Eq.~(\ref%
{Eq17}) in the following, we assume that $\omega _{ad_1}=\omega _{ad}$, $\omega
_{bd_2}=\omega _{bd}$ and $\omega _{cd_3}=\omega _{cd}$,
i.e., the three lower levels $|d_{j}\rangle $ ($j=1,2,3$) are degenerate, so that we have $\delta
_{k}^{\prime }=\delta _{k}$, $\omega _{ab}^{\prime }=\omega _{ab}$, $%
\omega _{ca}^{\prime }=\omega _{ca}$, $\delta
_{k}^{b\prime }=\delta^b _{k}$, $\delta
_{k}^{c\prime }=\delta^c _{k}$, $\gamma^{\prime }_a=\gamma_a$,  $\gamma^{\prime }_b=\gamma_b$, and  $\gamma^{\prime }_c=\gamma_c$. However, we should
point out that this assumption does not bring significant differences in
physical appearance except the positions of resonance peaks in the
spectra.

\end{widetext}

\section{Nonreciprocal transition and the spontaneous emission spectra}

\begin{figure}[tbp]
\includegraphics[bb=139 276 456 614, width=6 cm, clip]{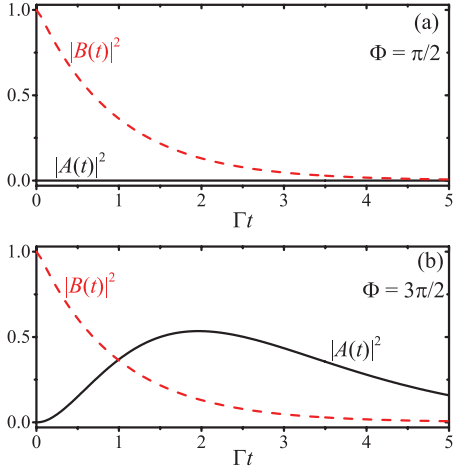}
\caption{(Color online) The populations $|A\left( t\right)|^2$ (black solid
curve) and $|B\left( t\right)|^2$ (red dashed curve) are plotted as
functions of the time $\Gamma t$ for: (a) $\Phi=\protect\pi/2$ and (b) $%
\protect\Phi=3\protect\pi/2$. The initial conditions are $A\left( 0\right)=C\left( 0\right)=0$
and $B\left( 0\right)=1$. The other parameters are $\protect\gamma_a=\protect%
\gamma_b=\Gamma/100$, $\protect\gamma_c=100\Gamma$, $\Omega_{ab}=\Gamma/2$, $%
\Omega_{ca}=\Omega_{bc}=5\Gamma$, $\Delta _{cb}=\Delta_{ca}=\Delta _{ab} =0$%
. }
\label{fig2}
\end{figure}

\begin{figure}[tbp]
\includegraphics[bb=114 315 470 607, width=8.5 cm, clip]{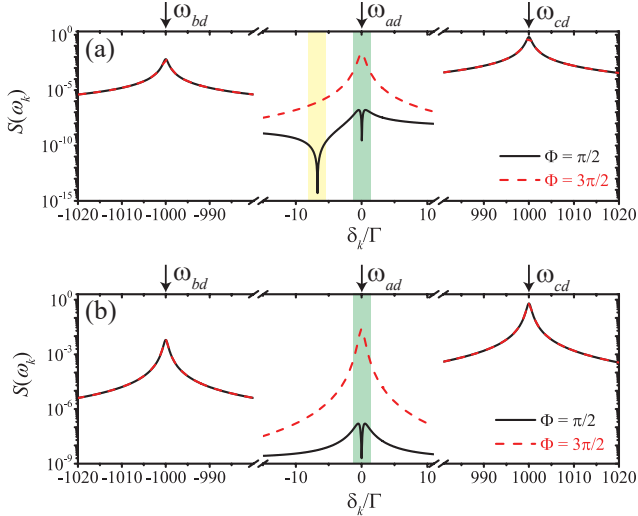}
\caption{(Color online) The spontaneous emission spectra for the three upper
levels ($|a\rangle$, $|b\rangle$, and $|c\rangle$) coupled by the same vacuum
modes to (a) the common lower level ($|d\rangle$), or (b) different lower
levels ($|d_1\rangle$, $|d_2\rangle$, and $|d_3\rangle$). The system is
initially in level $|b\rangle$. The black solid curves are for phase $\Phi=%
\protect\pi/2$ and the red dashed curves for $\Phi=3\protect\pi/2$. The
other parameters are $\protect\gamma_a=\protect\gamma_b=\Gamma/100$, $%
\protect\gamma_c=100\Gamma$, $\Omega_{ab}=\Gamma/2$, $\Omega_{ca}=%
\Omega_{bc}=5\Gamma$, $\Delta _{cb}=\Delta_{ca}=\Delta _{ab} =0$, and $%
\protect\omega_{ab}=\protect\omega_{ca}=10^3\Gamma$.}
\label{fig3}
\end{figure}

First, let us investigate the time evolution of the upper-level
populations with the system initially in level $|b\rangle$. The populations $%
|A\left( t\right)|^2$ (black solid curve) and $|B\left( t\right)|^2$ (red
dashed curve) obtained from Eqs.~(\ref{Eq10})-(\ref{Eq12}) are plotted as
functions of the time $t$ in Fig.~\ref{fig2}. The population can transfer
from the level $|b\rangle$ to level $|a\rangle$ for $\Phi=3\pi/2$, but
almost no population will transfer from the level $|b\rangle$ to level $%
|a\rangle$ when $\Phi=\pi/2$, which agrees with the result given in Ref.~%
\cite{XWXuArx19}. It has been shown in Ref.~\cite{XWXuArx19} that the transition probability for $|b\rangle\rightarrow |a\rangle $ is different from the one for $|a\rangle\rightarrow |b\rangle $ when $\Phi\neq n \pi$ ($n$ is an integer), i.e., the transitions between levels $|b\rangle$ and $
|a\rangle$ are nonreciprocal. As the
decay induced coupling terms with $p_{\sigma\sigma^{\prime }}\frac{\sqrt{\gamma _{\sigma}\gamma _{\sigma^{\prime }}}}{2}e^{\pm i\omega_{\sigma\sigma^{\prime }}t}$) in Eqs.~(\ref{Eq3})-(\ref{Eq5}) are high-frequency oscillating terms and neglected safely
under the assumption $\{\omega _{ab},\omega _{ca},\omega _{cb}\}\gg
\{\gamma_a,\gamma_b,\gamma_c,\Delta _{ab},\Delta _{ca},\Delta_{cb}\}$, the time evolutions of the upper-level populations
are the same for both the two cases with one common or three different lower levels.

The nonreciprocal transitions can be observed by measuring the spontaneous
emission spectra of the system. When $\Phi=3\pi/2$, as the population can
transfer from the level $|b\rangle$ to level $|a\rangle$, there should be a
peak around the resonant frequency $\omega_{ad}$. Instead, there is almost no
population transferring from the level $|b\rangle$ to level $|a\rangle$ when $%
\Phi=\pi/2$, so that the peak around the frequency $\omega_{ad}$ will be
eliminated (i.e., a dip should appear at there). In Fig.~\ref{fig3}, the
spontaneous emission spectra of the system are plotted for the three upper
levels ($|a\rangle$, $|b\rangle$, and $|c\rangle$) coupled by the same vacuum
modes to (a) the common lower level ($|d\rangle$), or (b) different lower
levels ($|d_1\rangle$, $|d_2\rangle$, and $|d_3\rangle$). The black solid
curves are for phase $\Phi=\pi/2$ and the red dashed curves for $\Phi=3\pi/2$%
. As expected for the spectra in both Figs.~\ref{fig3}(a) and \ref{fig3}(b),
there is a peak at the transition frequency $\omega_{ad}$ when $\Phi=3\pi/2$
or a dip when $\Phi=\pi/2$. Comparing the curves of Fig.~\ref{fig3}(a) and %
Fig.~\ref{fig3}(b), we find that there is another dip around $\delta_k/\Gamma=-6.7$ in the spectrum of the system with one
common lower level when $\Phi=\pi/2$, which is induced by destructive interference
between different decay paths to one common lower level~\cite%
{SYZhuPRA95,SYZhuPRL96}.

\section{Spontaneous emission spectra of chiral molecules}

\begin{figure}[tbp]
\includegraphics[bb=4 376 586 613, width=8.5 cm, clip]{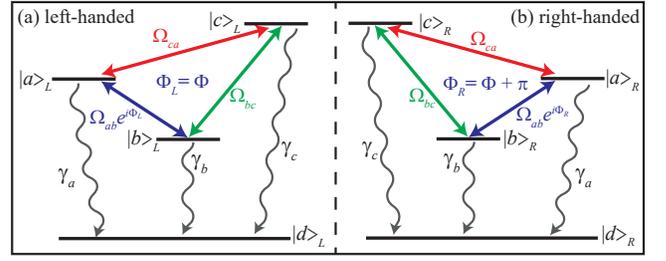}
\caption{(Color online) Level diagram of a chiral molecule with cyclic
transitions for the three upper levels ($|a\rangle_Q$, $|b\rangle_Q$ and $%
|c\rangle_Q$), and they are coupled by the same vacuum modes to the common
lower level ($|d\rangle_Q$): (a) $Q=L$ for left-handed chiral states and (b)
$Q=R$ for right-handed chiral states.}
\label{fig4}
\end{figure}

\begin{figure}[tbp]
\includegraphics[bb=139 343 465 592, width=8.5 cm, clip]{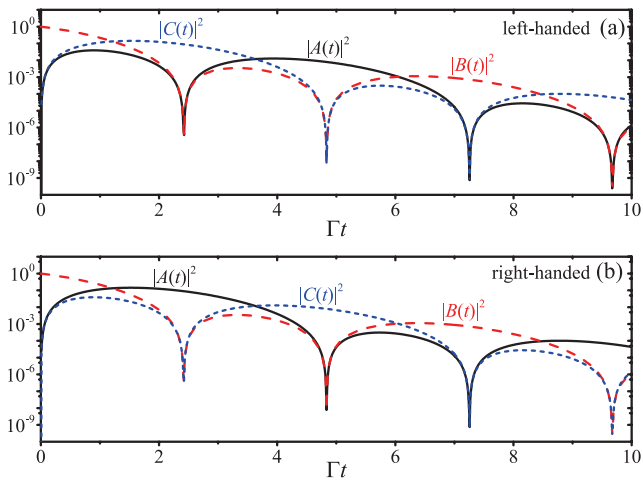}
\caption{(Color online) The populations $|A\left( t\right)|^2$ (black solid
curve), $|B\left( t\right)|^2$ (red dashed curve) and $|C\left( t\right)|^2$
(blue dot curve) are plotted as functions of the time $\Gamma t$ for: (a)
left-handed molecules and (b) right-handed molecules. The initial conditions
are $A\left( 0\right)=C\left( 0\right)=0$ and $B\left( 0\right)=1$. The
other parameters are $\protect\gamma_a=\protect\gamma_b=\protect\gamma%
_c=\Gamma$, $\Omega_{ab}=\Omega_{ca}=\Omega_{bc}=\Gamma/2$, $\Delta
_{cb}=\Delta_{ca}=\Delta _{ab} =0$, and $\Phi=\protect\pi/2$.}
\label{fig5}
\end{figure}

\begin{figure}[tbp]
\includegraphics[bb=115 343 467 609, width=8.5 cm, clip]{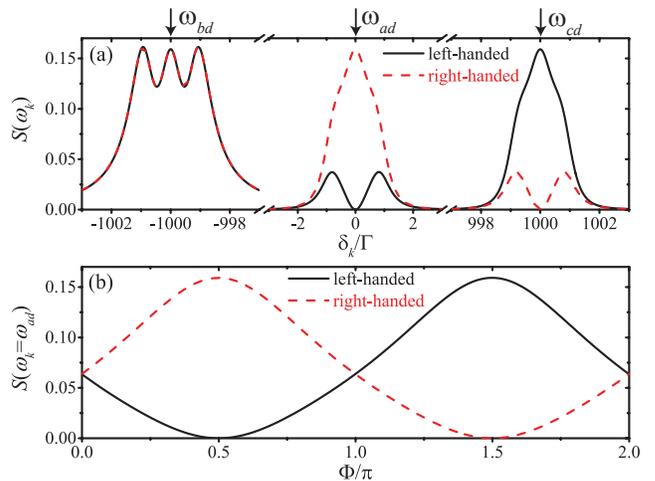}
\caption{(Color online) The spontaneous emission spectra for a (left-handed or right-handed) chiral
molecule with three upper levels ($|a\rangle_Q$, $|b\rangle_Q$ and $%
|c\rangle_Q$) coupled by the same vacuum modes to a common lower level ($%
|d\rangle_Q$). The molecule is initially in level $|b\rangle_Q$. The black solid
curves are for the ($Q=L$) left-handed molecule and the red dashed curves for the ($%
Q=R$) right-handed molecule. (a) The spontaneous emission spectra versus $%
\protect\delta_k$ for $\Phi=\protect\pi/2$; (b) the spontaneous emission
spectra versus $\Phi$ for $\protect\omega_k=\protect\omega_{ad}$. The other
parameters are $\protect\gamma_a=\protect\gamma_b=\protect\gamma_c=\Gamma$, $%
\Omega_{ab}=\Omega_{ca}=\Omega_{bc}=\Gamma/2$, $\Delta
_{cb}=\Delta_{ca}=\Delta _{ab} =0$, and $\protect\omega_{ab}=\protect\omega%
_{ca}=10^3\Gamma$.}
\label{fig6}
\end{figure}

As an important application, we will discuss how to realize the
determination of enantiomeric excess based on the spectral line elimination
in the spontaneous emission spectra of chiral molecules. Our method is based
on the model of chiral molecules with cyclic-transition three upper levels
coupled with one common lower level by the same vacuum modes, as shown in
Fig.~\ref{fig4}, where $|a\rangle_Q $, $|b\rangle_Q $, $|c\rangle_Q $, and $%
|d\rangle_Q $ ($Q=L,R$) are the inner states of the left- and right-handed
molecules. Both the models of left- and right-handed molecules can be described by the Hamiltonian given in Eq.~(\ref%
{Eq1}) with $\Phi=\Phi_L$ for left-handed molecules and $\Phi=\Phi_R-\pi$
for right-handed molecules.

Let the chiral molecule initially be in level $|b\rangle_Q$. The time
evolution of the populations in the three upper levels is shown in Fig.~\ref%
{fig5}. For the left-handed molecule, the population is transferred from $%
|b\rangle_L$ to $|c\rangle_L$ first, then to $|a\rangle_L$, and last back to
$|b\rangle_L$, with the maximum population decaying exponentially. Conversely,
the population of the right-handed molecule is transferred from $|b\rangle_L$
to $|a\rangle_L$ first, then to $|c\rangle_L$, and last back to $|b\rangle_L$%
, with the maximum population decaying exponentially also. The
chirality-dependent population transferring can be used for inner-state
enantio-separation~\cite%
{KralPRL01,KralPRL03,YLiPRA08,WZJiaJPB10,LehmannJCP18,VitanovPRL19,CYePRA19,CYearX19,JLWuarX19}. We note that
the similar cyclic transitions between three levels have been observed in a
ring with three transmon superconducting qubits~\cite{RoushanNPy17} and a
single spin under closed-contour interaction~\cite{BarfussNPy18}.
Nevertheless, the exponential decay of the maximum population is one
ingredient for nonreciprocal transition between the three upper levels, and
the direction of the transition is chirality-dependent. In Fig.~\ref{fig5},
there is much more population transferred from $|b\rangle_L$ to $|c\rangle_L$
than that from $|b\rangle_L$ to $|a\rangle_L$ for left-handed molecules, and much more population
transferred from $|b\rangle_L$ to $|a\rangle_L$ than that from $|b\rangle_L$ to $|c\rangle_L$ for
right-handed molecules.

The nonreciprocal transition in the chiral molecules can be reflected in the
spontaneous emission spectra. The spontaneous emission spectra for chiral
molecules are plotted as a function of detuning $\delta_k$ in Fig.~\ref{fig6}(a)
with the black solid curves for left-handed molecules and the red dashed
curves for right-handed molecules when $\Phi=\pi/2$. The most obvious
difference between the two curves is there is one peak (with value $S_L$) at transition frequency $%
\omega_{cd}$ for left-handed molecules, but the peak is eliminated and a
dip appears for right-handed molecules. By contrast, there is one peak (with value $S_R$) at transition
frequency $\omega_{ad}$ for right-handed molecules, but the peak is
eliminated for left-handed molecules.
Then, the strengths of the
spontaneous emission spectra of a chiral mixture, $S_M(\omega)$, at frequencies $\omega_{cd}$ and $\omega_{ad}$
are proportional to the molecule numbers of the two enantiomers, respectively,
\begin{equation}
S_M(\omega_{cd})\approx N_L S_L,
\end{equation}
\begin{equation}
S_M(\omega_{ad})\approx N_R S_R,
\end{equation}
where $N_L$ ($N_R$) is the number of left-handed (right-handed) molecules.
For the enantiomeric excess of a chiral mixture defined by $\varepsilon \equiv (N_L-N_R)/(N_L+N_R)$, the enantiomeric excess can be determined by
\begin{equation}
\varepsilon=\frac{S_M(\omega_{cd})-\eta S_M(\omega_{ad})}{S_M(\omega_{cd})+\eta S_M(\omega_{ad})}
\end{equation}
with the coefficient $\eta\equiv S_L/S_R$. Here, we have $\eta \approx 1$ with the parameters used in Figs.~\ref{fig5} and \ref{fig6}.
In addition, the spontaneous emission spectra for chiral
molecules show a sinusoidal dependence on the phase $\Phi$, as shown in Fig.~%
\ref{fig6}(b). $\Phi=\pi/2$ and $\Phi=3\pi/2$ are two optimal phases to make
the determination of enantiomeric excess mostly efficiently.

\section{Conclusions}

In conclusion, we have studied the spontaneous emission spectrum of a
multi-level system with nonreciprocal transition. Spectral line elimination
appears in the spectra when there is almost no population transferring in one
of the transition directions for nonreciprocal transition. The spectral line
elimination induced by nonreciprocal transition provides us a method to
determine the enantiomeric excess for the chiral molecules without requiring the enantio-pure
samples. When the spectral line
elimination appears at some resonant frequencies for molecules with one
chirality, the strengths of the spontaneous emission spectra at that
frequencies are proportional to the numbers of the molecules with
the opposite chirality, so the enantiomeric excess can be determined by
measuring the strengths of the spontaneous emission spectra at these
frequencies.

\vskip 2pc \leftline{\bf Acknowledgement}

X.-W.X. is supported by the National Natural Science Foundation of China
(NSFC) under Grant No.~11604096, and the Key Program of Natural Science
Foundation of Jiangxi Province, China under Grant No.~20192ACB21002. Y.L.
is supported by National Key R\&D Program of China under Grant
No.~2016YFA0301200, and NSFC under Grants No.~11774024, No.~U1930402, and
No.~U1730449. A.-X.C. is supported by NSFC under Grant No.~11775190.

\end{document}